\documentstyle[sprocl]{article}

\input{psfig}
\def\be{\begin{equation}}
\def\ee{\end{equation}}
\def\bea{\begin{eqnarray}}
\def\eea{\end{eqnarray}}

\newcommand{\pom}{I\!\! P}
\begin{document}

\title{POMERON: BEYOND THE STANDARD APPROACH
\footnote{To appear in Proceedings of 
``XXIX International Symposium on Multiparticle Dynamics, 
9-13 August 1999, Brown University, Providence, RI 02912, USA".}
}
\author{K. GOULIANOS}

\address{The Rockefeller University, 1230 York Avenue, New York,\\ 
NY 10021, USA\\E-mail: dino@physics.rockefeller.edu}




\maketitle\abstracts{
We discuss the experimental evidence supporting the concept of 
universality of the rapidity gap probability in soft and hard 
diffraction, relate the gap probability to hadronic parton 
densities, and present a phenomenological model of diffraction
in which the structure of the Pomeron is derived from the 
structure of the parent hadron. Predictions for diffractive 
deep inelastic scattering are compared with data.
}
  
\section{Introduction}
Experiments at HERA and at hadron colliders have 
reported and characterized a class of events incorporating a hard 
(high transverse momentum) partonic scattering while carrying the 
characteristic 
signature of diffraction, namely a leading (anti)proton and/or 
a large {\em rapidity gap} (region of pseudorapidity
devoid of particles).
The prevailing theoretical idea is that the 
exchange across the gap
is the Pomeron~\cite{Regge}, which 
in QCD is a color-singlet exchange of gluons and/or quarks
with vacuum quantum numbers. 

A question of intense theoretical
debate is whether
the Pomeron has a unique particle-like partonic structure.
This question can be addressed experimentally by 
comparing the parton distribution functions ({\em pdf'\,}s) 
of the (anti)proton measured in a variety of hard single diffraction 
dissociation processes as a function of $\xi$, $t$, $Q^2$ and $x$
(or $\beta\equiv x/\xi$), where $t$ is the 4-momentum transfer and $\xi$ 
the fractional momentum loss of the (anti)proton.
The gluon and quark {\em pdf'\,}s can be sorted 
out by studying processes  with different sensitivity 
to the gluon and quark components of the Pomeron. 

The proton diffractive {\em pdf's\,} 
have been measured in diffractive deep inelastic scattering (DDIS) 
at HERA by both the 
H1~\cite{H1} and ZEUS~\cite{ZEUS} Collaborations. 
The experiments measure directly the $F_2$ diffractive
structure function, $F_2^{D(3)}(\xi,Q^2,\beta)=
\int_{t=-1}^{t_{min}}F_2^{D(4)}(t,\xi,Q^2,\beta)dt$. The variable 
$\xi$ is related to the rapidity gap 
by $\Delta y=\ln\frac{1}{\xi}$.   
The gluon diffractive $pdf$ was determined by H1 from a QCD analysis of the 
$Q^2$ evolution of $F_2^{D(3)}$.
All HERA hard diffraction results are generally consistent with the 
parton densities obtained from DDIS.   
However, assuming factorization, calculations based on these $pdf$'s 
predict~\cite{KGDIS,ACTW,CS} rates $\sim 10$ times
larger than the
measured $W$ and dijet production rates at the Tevatron~\cite{CDF}. 
The suppression
of Tevatron relative to HERA diffractive rates represents a breakdown of 
factorization.

The magnitude of the factorization breakdown is 
in general agreement with predictions based on the renormalized Pomeron 
flux model~\cite{R}, in which the Pomeron flux (see next section)
is viewed~\cite{lishepsum} as a rapidity gap probability density and 
is normalized by scaling it to its integral 
over the available phase space in $(\xi,\,t)$. In addition to its success 
in predicting soft and hard diffraction rates, 
the flux renormalization model describes 
{\em differential} soft diffraction cross sections~\cite{GM}
remarkably well.
However, the $\beta$-dependence of the Pomeron/diffractive $pdf$'s 
is not specified by the model
and had to be introduced ``by hand". In this paper, 
the diffractive $pdf$'s are derived from the 
non-diffractive using a parton model approach to hard diffraction 
based on the the concept of a normalized 
rapidity gap probability. 

\section{Clues from soft physics}
The Pomeron was introduced in Regge theory to 
account for the high energy behavior of the elastic, diffractive and total 
hadronic cross sections. In terms of the Pomeron trajectory,
$\alpha(t)=1+\epsilon+\alpha' t$, the $pp$ cross sections can be written as
{\large
\begin{equation}
\sigma_T(s)=\beta^2_{\pom pp}(0)
\left(\frac{s}{s_0}\right)^{\alpha(0)-1}
\sim {\left(\frac{s}{s_0}\right)}^{\epsilon}
\label{total}
\end{equation}
\begin{equation}
\frac{d\sigma_{el}}{dt}=\frac{\beta^4_{\pom pp}(t)}{16\pi}\;
{\left(\frac{s}{s_0}\right)}^{2[\alpha(t)-1]}
\sim \exp[(b_{\circ}+2\alpha'\ln{s}) t]
\label{elastic}
\end{equation}
\begin{equation}
\frac{d^2\sigma_{sdd}}{d\xi dt}=
\underbrace{\frac{{\beta_{\pom pp}^2(t)}}{16\pi}\;
\frac{1}{\xi^{2\alpha(t)-1}}}_
{{\rm Pomeron\;flux}\,\sim \frac{1}{\xi^{1+2\epsilon}}}\;
\underbrace{\beta_{\pom pp}(0)\,g(t)
\;\left(\frac{\xi s}{s_0}\right)^{\alpha(0)-1}}_
{\sigma_T^{\pom p}\sim (\xi s)^{\epsilon}}
\label{diffractive}
\end{equation}
}
The Regge approach has been successful in predicting 
the three salient features 
of high energy behavior, namely  (i) the rise of the total cross 
secions with energy~\cite{DL,CMG}, (ii) the shrinking of the forward 
elastic scattering peak~\cite{KG}, and (iii) the shape of the 
$M^2$ dependence of SDD~\cite{GM,KG}. However, these features are 
also present in a parton model approach~\cite{Levin}, as outlined below.
\subsection{Rise of total cross sections}
The $pp$ total cross section
is basically proportional to the number of partons in the proton, 
integrated down to $x=s_{\circ}/s$, where $s_{\circ}$ is the 
energy scale for soft physics.  
The latter is of ${\cal{O}}(\langle M_T\rangle^2)$, 
where $\langle M_T\rangle\sim 1$ GeV 
is the average transverse mass of the particles in the final state.
Expressing the parton density as a power law in $1/x$, which is an appropriate 
parameterization for 
the small $x$ region responsible for the cross section rise at
high energies, we obtain
{\large 
\begin{equation}
\sigma_T\sim
{\displaystyle\int}_{(s_{\circ}/s)}^1\;\frac{dx}{x^{1+n}}
\sim\left(\frac{s}{s_{\circ}}\right)^{n}
\end{equation}
}
The parameter $n$ is identified with the 
$\epsilon=\alpha(0)-1$ of the Pomeron trajectory. 
\subsection{Shrinking of forward elastic peak}
In Regge theory, the slope of the forward elastic peak increases as $\ln s$,
while in the parton picture one would naively expect the slope 
to follow the increase of the total cross section:
{\large 
\begin{equation}
\frac{d\sigma^{el}}{dt}
\sim \underbrace{\exp[(b_{\circ}+2\alpha'\ln{s}) t]}_{\rm Regge}
\Rightarrow 
\underbrace{\sim \exp[(b_{\circ}+cs^{n}) t]}_
{\rm Parton\;model}
\end{equation}
}
Using $n=\epsilon=0.104$~\cite{GM}, 
we note that from $\sqrt{s}=20$ to 1800 GeV 
the terms $\ln s$ and $s^{n}$ increase by factors of 2.50 and 
2.55, respectively. Thus,  the Regge and parton model expressions 
for the slope are experimentally indistinguishable. 
\subsection{$M^2$ dependence of diffraction dissociation}
The diffractive mass is related to $\xi$ by $M^2\approx \xi s$.
In the Regge picture, the $t=0$ single diffractive 
cross section~\footnote{In the rest of 
this paper we use $t=0$ cross sections for simplicity, but the results we 
obtain generally apply also to the cross sections integrated over $t$.}
is given by (see Eq.~\ref{diffractive})
{\large 
\begin{equation}
{\rm Regge:}\;\;\;\frac{d\sigma_{sdd}}{dM^2}
\sim \frac{s^{2\epsilon}}{(M^2)^{1+\epsilon}}
\end{equation}
}
However, experimentally the cross section is found to be independent 
of $s$~\cite{R,GM}:
{\large 
\begin{equation}
{\rm Experiment:}\;\;\;\frac{d\sigma_{sdd}}{dM^2}
\sim \frac{1}{(M^2)^{1+\epsilon}}
\label{scaling}
\end{equation}
}
In terms of $\xi=\frac{M^2}{s}$, we have 
$\frac{d\sigma^{\rm exp}_{sdd}}{d\xi}\sim \frac{1}{s^{\epsilon}}\;
\frac{1}{\xi^{1+\epsilon}}$,
which can be written as
{\large
\begin{equation}
\frac{d\sigma_{sdd}}{d\xi}
\sim \underbrace{\frac{1}{s^{2\epsilon}}\;
\frac{1}{\xi^{1+2\epsilon}}}_{}\times (\xi s)^{\epsilon}
\label{sdd}
\end{equation}
}
Recalling that $\xi$ is related to the rapidity gap and 
$\xi s=M^2$ is the s-value of the diffractive sub-system, and noting that 
${\int}_{(s_{\circ}/s)}^{1}\;
\frac{1}{s^{2\epsilon}}\;\frac{d\xi}{\xi^{1+2\epsilon}}\,=\,{\rm constant}$,
Eq.~\ref{sdd} may be viewed as representing
the product of the total cross section at the sub-system energy 
multiplied by a {\em normalized} rapidity gap probability. 
This {\em experimentally established} scaling behavior 
is used in the next section 
as a clue in deriving the diffractive $F_2$ structure function 
from the non-diffractive in the parton model approach to diffraction. 

\section{Diffractive deep inelastic scattering}
The inclusive and diffractive DIS cross sections are proportional to 
the corresponding $F_2$ structure functions of the proton, 
{\large
\begin{equation}
\frac{d^2\sigma}{dxdQ^2}\sim \frac{F^h_2(x,Q^2)}{x};
\;\;\;\;\;\;\;
\frac{d^3\sigma}{d\xi dxdQ^2}\sim \frac{F^{D(3)}_2(\xi,x,Q^2)}{x} 
\label{SF}
\end{equation}
}
\noindent where the superscripts $h$ and $D(3)$ indicate 
a {\em hard} inclusive structure (at the scale $Q^2$) and a 3-variable
diffractive structure (integrated over $t$). The latter depends not only 
on the hard scale $Q^2$, but also on the {\em soft} scale 
$\langle M_T\rangle^2$, which is the relevant one for the 
formation of the gap. 

The only marker of the rapidity gap is the variable $\xi$. We therefore 
postulate that the rapidity gap probability is proportional to the 
{\em soft} parton density at $\xi$ and write the DDIS cross section as
{\large 
\begin{equation}
\frac{d^3\sigma}{d\xi dxdQ^2}\sim \frac{F^h_2(x,Q^2)}{x} \times
\frac{F^s_2(\xi)}{\xi}\otimes \xi-{\rm norm}
\end{equation}
}
\noindent where the symbolic notation ``$\otimes \;\xi-{\rm norm}$" is 
used to indicate that the $\xi$ probability is normalized.
Since $x=\beta\xi$, the normalization over all available $\xi$
values involves not only $F_2^s$ but also $F_2^h$, breaking down factorization. 
It is therefore prudent to write the DDIS cross section in terms of $\beta$
instead of $x$, so that the dependence of $F_2^h$ on $\xi$ is shown 
explicitely: 
{\large 
\begin{equation}
\frac{d^3\sigma}{d\xi d\beta dQ^2}\sim
\frac{F^h_2(\beta\xi,Q^2)}{\beta} \times
\frac{F^s_2(\xi)}{\xi}\otimes \xi-{\rm norm}
\end{equation}
}
\noindent The term to the right of the $\sim$ sign represents the 
$F_2^{D(3)}(\xi,\beta,Q^2)$ 
structure function. Guided by the scaling behaviour of Eq.~\ref{sdd}, we now
factorize $F_2^{D(3)}(\xi,\beta,Q^2)$ into $F_2^h(\beta,Q^2)$, 
the sub-energy cross section, times a normalized gap probability:
{\large 
\begin{equation}
F_2^{D(3)}(\xi,\beta,Q^2)=
P_{gap}(\xi,\beta,Q^2)
\times F^h_2(\beta,Q^2)
\label{F2D3}
\end{equation}
}
The gap probability is given by
{\large 
\begin{equation}
P_{gap}(\xi,\beta,Q^2)=
\frac{F^h_2(\beta\xi,Q^2)}{\beta}\times \frac{F^s_2(\xi)}{\xi}
\times N(s,\beta,Q^2)
\label{GP}
\end{equation}
}
\noindent where $N(s,\beta,Q^2)$ is the normalization factor, 
{\large 
\begin{equation}
N(s,\beta,Q^2)=f_q\;/\;{\displaystyle\int}_{\xi_{min}=Q^2/\beta s}^1
\frac{F^h_2(\beta\xi,Q^2)}{\beta}\times \frac{F^s_2(\xi)}{\xi} d\xi
\label{GN}
\end{equation}
}
\noindent in which $f_q$ denotes the quark fraction of the 
hard structure (since only quarks participate in DIS). 

\subsection{Comparison with data}
At small $x$ and small $\xi$, the structure functions $F_2^h$ and $F_2^s$ 
are represented well by
$F^h_2(x,Q^2)={A^h}/{x^{\epsilon^h(Q^2)}}$ and 
$F^s_2(\xi)={A^s}/{\xi^{\epsilon^s}}$. An analytic evaluation of 
Eqs.~\ref{GN} and \ref{F2D3} yields
{\large
\begin{equation}
N(s,\beta,Q^2)=f_q\;/\;\left[
\frac{A^h}{\beta^{1+\epsilon^h}}\;\frac{A^s}{\epsilon^h+\epsilon^s}\;
\left(\frac{\beta s}{Q^2}\right)^{\epsilon^h+\epsilon^s}\right]
\end{equation}
}
{\large 
\begin{equation}
F_2^{D(3)}(\xi,\beta,Q^2)=
\frac{1}{\xi^{1+\epsilon^h+\epsilon^s}}\;
\times f_q(\epsilon^h+\epsilon^s)\;\left(\frac{Q^2}{\beta s}\right)^
{\epsilon^h+\epsilon^s}
\times \frac{A^h}{\beta^{\epsilon^h}}
\label{F2D3A}
\end{equation}
}
\newpage
\noindent \begin{minipage}[t]{6cm}
Figure~1 shows a comparison of the $\beta$ distribution of H1 data~\cite{H1}
at $\xi\approx 0.01$ and $Q^2=45$ GeV$^2$
with the prediction of Eq.~\ref{F2D3A}. 
The following parameters were used in the calculation:
$\sqrt{s}=280$ GeV, $\epsilon^s=0.1$~\cite{CMG}, $Q^2=45$ GeV$^2$, 
$\epsilon^h(Q^2=45)=0.3$, $f_q=0.4$, $\xi=0.01$ and $A^h=0.2$, evaluated from 
$F_2(Q^2=50,x=0.00133)=1.46\Rightarrow\frac{A^h}{x^{0.3}}$.
The agreement between theory and experiment may be considered excellent, 
particularly since no free parameters were used in the calculation.
\end{minipage}
\hfill
\noindent\begin{minipage}[t]{5cm}
\vfill
\vglue -0.9cm
\centerline{\psfig{figure=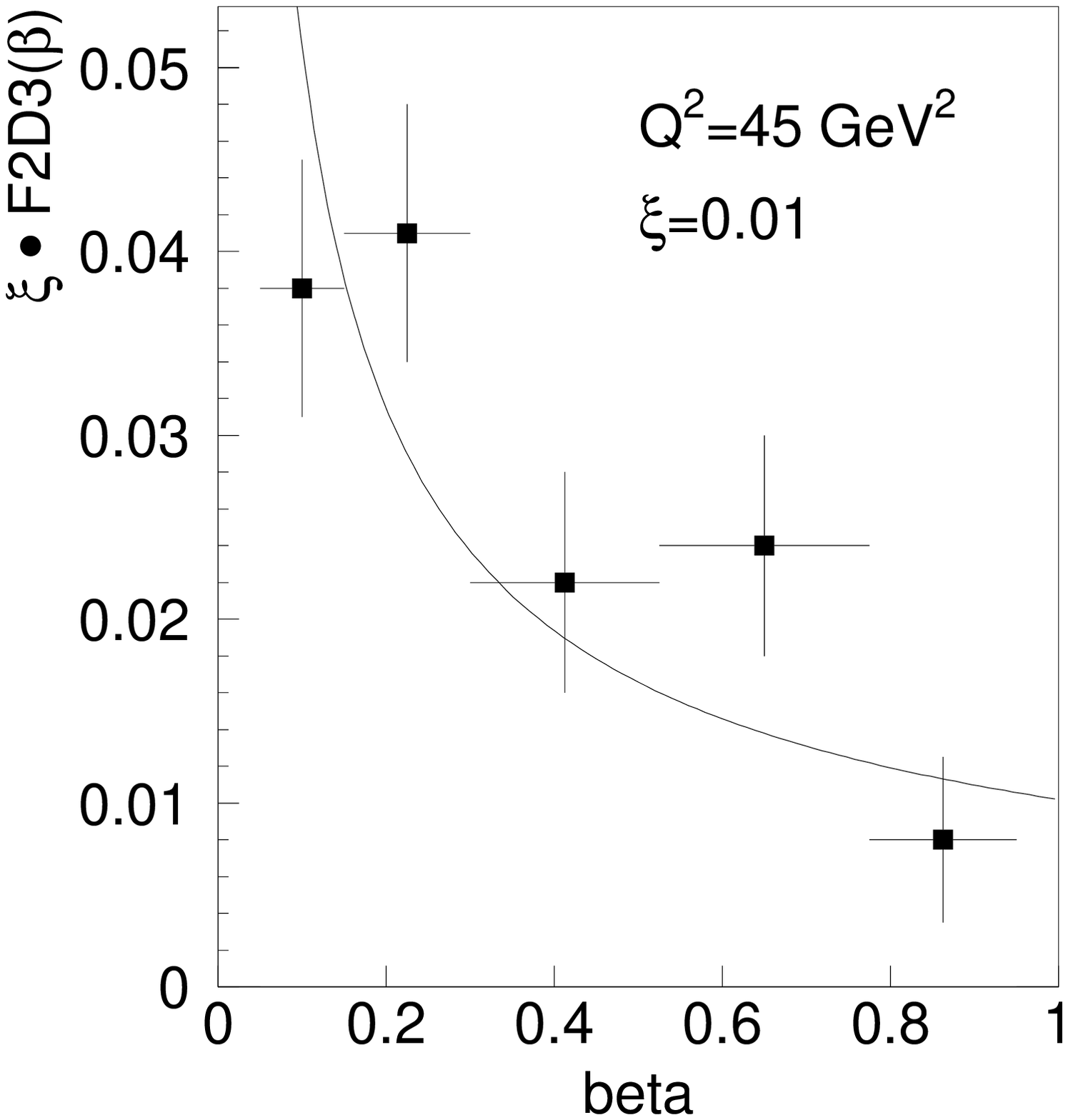,width=2.5in}}
\centerline{Figure 1}
\end{minipage}

\end{document}